\documentclass[runningheads]{llncs}
\usepackage[T1]{fontenc}
\usepackage{graphicx}
\usepackage{todonotes}
\usepackage{hyperref}
\usepackage{color}

\begin{document}
\title{IN2P3 Computing Center 2024 Workload Dataset}
\author{Guillaume Cochard\inst{1} \and Bertrand Simon\inst{1,2}}
\authorrunning{G. Cochard and B. Simon}
\institute{CC-IN2P3, CNRS, Villeurbanne, France\\
\email{guillaume.cochard@cc.in2p3.fr}\\ \and
Université Grenoble Alpes, CNRS, INRIA, Grenoble INP, LIG, Grenoble, France\\
\email{bertrand.simon@cnrs.fr}}
\maketitle            
\begin{abstract}
This paper provides and analyzes a dataset detailing the characteristics and execution data of all jobs submitted to the IN2P3 Computing Center (Villeurbanne, France), a national research and support unit of the CNRS, in 2024. The main additional value of this contribution compared to previously available datasets consists in the combination of an extended time interval considered, the inclusion of memory usage data and its recency, on top on improving the diversity of datasets provenance.
This allows researchers to simulate and evaluate scheduling algorithms on a real workload over a large time window. Thus, specificities due to seasonal, monthly, and weekly user behaviors can be taken into account, which is not possible with smaller or synthetic datasets. It is composed of 44M jobs submitted by 1k users running on a cluster of a maximum of 312 machines supporting 46k concurrent threads and providing 105To of RAM.

\end{abstract}

\section{Introduction}
\label{sec:intro}

Many research projects on scheduling need real data to evaluate and validate the solutions developed. Therefore, several supercomputers or computing centers have dedicated some effort to create public datasets describing the jobs submitted on their system. Nevertheless, finding appropriate traces with enough information collected in order to analyze the desired behavior remains a challenging task. Available datasets may not span a time window long enough to exhibit relevant submission patterns, may not disclose required information such as the requested time limit or the memory usage of the jobs, may be composed of old outdated job characteristics, or may not be diverse enough to avoid overfitting behaviors. Researchers may therefore resort to using old data, see~\cite{haddad2021stand,haddad2021combined,vasconcelos2022indirect,villebonnet2016energy} or synthetically generated datasets to analyze the algorithms developed, see for instance~\cite{benaissa2022standalone,vasconcelos2023optimal,cendrier2025green}.

The objective of this paper is to provide and analyze a new workload dataset\footnote{\url{https://zenodo.org/records/18668107}} complementing  the publicly available collection of job traces. It is composed of jobs submitted to the IN2P3 (National Institute of Nuclear and Particle Physics) Computing Center (CC-IN2P3)\footnote{\url{https://cc.in2p3.fr/en/qui-sommes-nous/le-cc-in2p3/}},  a national research and support unit of the CNRS (National Center for Scientific Research) located in Villeurbanne, France. It spans a long time-window of one entire year, 2024, allowing to analyze algorithms without ignoring user behaviors dependence on the current day, week or month. Multiple descriptors are provided for each job, collected mainly through the SLURM scheduler~\cite{slurm} system. Submission characteristics include user identification, partition requested, submission time, and requested timelimit, memory and CPU count. Post-execution information regarding each job also include eligible, start and end times, total CPU time, maximum resident set size (peak memory used), amount of data written to disk and machine identifier. We also provide details on the cluster on which these jobs have been submitted, in order to improve reproducibility.

The exact computing capabilities of the cluster evolve, and attain at the end of the year 312 machines able to run 46k concurrent threads and offer 105To of RAM. In total, the resources available over this year sum to approximately 179M CPU.hours and 677M of Gb.hours of RAM.

This allows to evaluate algorithms using real user demands and actual usage of resources (CPU, time, memory and disk writes) over a long time window. Jobs can also be classified based on their associated user, account or partition.

The rest of the paper is organized as follows. In Section~\ref{sec:cc}, we provide a general description of the computing center. In Section~\ref{sec:related}, we describe related available datasets. In Section~\ref{sec:dataset}, we explicit this dataset structure. In Section~\ref{sec:cluster}, we describe the scheduling system. The limitations are discussed in Section~\ref{sec:caveats}. Finally, in Section~\ref{sec:analysis}, we provide statistical descriptions of the dataset content.

\section{CC-IN2P3 description}
\label{sec:cc}

\subsection{Usage of the cluster}

\label{sec:usage}

The users of this cluster belong to the High-Energy Physics and Astrophysics
communities. It is designed to handle numerous relatively small jobs (few minutes to
few days), which are usually sent by batches of many jobs, hence the waiting time or execution
speed actually are secondary objectives. The main objective is to maximize the total computing power
available or allocated over a given period of time (e.g., a year), while supporting peak demands. 
Most jobs are submitted via human intervention from users, within semi-automated workflows or not, therefore the load varies depending on the current day and time.
In these communities, jobs used to request on average 3GB of memory per CPU, but this quantity has been recently increasing significantly. Consequently, high-load periods typically saturate the memory available within the cluster more often than the number of cores.
Typical jobs also request a single or few CPUs, but there is now a small portion of highly parallel or GPU-intensive jobs, hence a fraction of the cluster is dedicated to these needs.

The system underwent maintenance during four days (03-12, 06-25, 09-17, 12-03). Users had been warned ahead of time, and the currently running jobs got killed and queued back at the end of the maintenance. Note that all systems may not become fully operational again simultaneously.

Focusing on the main partition (\texttt{htc}, see Section~\ref{sec:partition}), a total of 69\% of available total CPU time has been allocated to jobs, while 41\% has been effectively used. Regarding memory usage, 95\% has been allocated while 50\% has been used by jobs. In order to avoid memory waste, a memory overbooking mechanism is implemented, so that, for each \texttt{htc} machine, the total memory that can be allocated is slightly larger than the actual available memory. As users typically overestimate their need, and job memory peaks are not simultaneous, the memory is rarely overutilised. In such a case, excess jobs are killed and resumed later. The overbooking averaged around 30\% over the whole time window, so approximately 70\% of the perceived memory has been allocated and 40\% used.

\subsection{Hardware composition}

The cluster is composed of a few hundreds of machines, thirty of them being added during the time window considered.

In the dataset, the machines denomination follows the naming convention \texttt{x01} where \texttt{x} is a letter denoting the machine type, as depicted in Table~\ref{tab:hardware}. Each machine belongs to one or several partitions, as described in Section~\ref{sec:partition}.
 Apart from \texttt{p[01-16]}, all those machines are hyperthreaded: each physical core is associated to two CPUs on which Slurm can allocate jobs.

The machines that have been added to the cluster during this year are  \texttt{d[01-28]} on 2024-10-04, \texttt{g13} on 2024-05-10 and \texttt{g14} on 2024-05-15.

\begin{table}[htbp]
	\centering
	\caption{Resources by machine types. IXS refers to Intel Xeon Silver, IXC to Intel Xeon CPU and AE to AMD EPYC.}
\label{tab:hardware}
\begin{tabular}{cccccc}
	\hline
	\textbf{Name} & \textbf{Nodes} & ~\textbf{CPUs}~ & \textbf{Memory} & \textbf{CPU type} & ~~\textbf{GPUs} \\
	\hline
	a[01-192] & 192 & 64 & 192 GB & AE 7302 16-Core Processor &  \\
	b[01-56] & 56 & 112 & 386 GB & AE 7453 28-Core Processor &    \\
	d[01-28] & 28 & 128 & 1289 GB & AE 9334 32-Core Processor &    \\
	g[02,10-14] & 6 & 20 & 192  GB& IXS 4114 CPU @ 2.20GHz & 4  v100 \\
	g[04-07,09,15-19] & 10 & 24 & 192 GB & IXS 4214R CPU @ 2.40GHz & 4  v100 \\
	g[01,08] & 2 & 24 & 1546 GB & IXS 4214R CPU @ 2.40GHz & 4  v100 \\
	m01 & 1 & 40 & 1546 GB & IXS 4114 CPU @ 2.20GHz &    \\
	p[01-16] & 16 & 32 & 128 GB & IXC E5-2698 v3 @ 2.30GHz &    \\
	\hline
\end{tabular}
\end{table}

\subsection{Storage and distributed filesystems}
The cluster operates three distributed filesystems, stored on disks and magnetic tapes depending on the volume and access frequency of files. Jobs usually mainly read and write from/to those filesystems, but may also use a local temporary space shared between jobs (jobs can only access files they own). This may largely improve the performance of I/O-bound jobs, but there is no usage limit or availability guarantee on this shared space.

\section{Related datasets}

\label{sec:related}

Several datasets similar to the one presented here have already been published. The main additional value of our contribution relies in the combination of a long time window, comprehensive job details (e.g., memory usage), and recency. It also improves the diversity of the available data. We report below an overview of the datasets currently publicly available, without claiming exhaustivity.

The Parallel Workload Archive~\cite{pwa} hosts 40 datasets from 1993 to 2018. They propose the Standard Workload Format~\cite{swf} as a standard to describe such logs. We have adopted a similar format with additional fields to provide information on the memory usage for instance, which is not accepted by the original format.
Similarly, the Grid Workload Archive~\cite{gwa} hosts 13 computing grid datasets up to 2013 but mostly from 2005-2007. They also propose an extension of the Standard Workload Format to accommodate additional relevant information.
The Failure Trace Archive~\cite{fta} is a complementary dataset that reports failures of nodes instead of focusing on jobs.
The SWIM project~\cite{swim} focuses on 2009-2011 MapReduce workloads from the companies Cloudera and Facebook. They focus on memory usage, and more specifically on input, shuffle, and output data size.
The Google cluster Borg~\cite{google} released one month of job traces from both years 2011 and 2019. As the industrial context is different from ours, the users do not need to submit walltime or memory limits.
A cluster from the company Alibaba~\cite{alibaba} released several datasets on specific fields such as AI-as-a-service. The dataset closest to the type of data we furnish covers an 8-day time window from 2018.
Aiming at diversifying the datasets available, especially because of the predominancy of the Google dataset, \cite{amvrosiadis2018diversity} propose two datasets, 9 months from a private cluster collected in 2016 and 5 years from the academic Mustang cluster collected in 2011-2016. These traces however do not include details on the memory consumption of jobs.
The most comprehensive dataset published covers operations from 2020-2022 of the Tier-0 supercomputer Marconi100~\cite{marconi}, totaling 50TB and including rare data such as hardware sensor monitoring.

\section{Description of the dataset}

\label{sec:dataset}

The dataset contains all jobs ended in 2024 (including some submitted in 2023) or submitted in 2024 (including some ended in 2025).
Sensitive data has been obfuscated or anonymized following recognized practices~\cite{obfuscation}.

\newcommand{\mykey}[2]{\texttt{#1:} #2}

The dataset is composed of 12 files, one per month of 2024, each representing a compressed CSV file with one line per job. Most fields have been obtained using the SLURM command  \texttt{sacct} \footnote{\url{https://slurm.schedmd.com/sacct.html}}, more information can be found in the official documentation. We provide two additional fields, \texttt{elapsed} and \texttt{el9}, to directly provide the duration of the job, and a custom flag explained below.

\begin{itemize}
	\item \mykey{jobid}{identifier of the job, sorted by submission time.}
	\item \mykey{state}{output state of the job, among \texttt{\{completed, timeout, canceled, failed, out\_of\_memory, node\_fail\}}}.
	\item \mykey{account}{unique anonymized identifier of the group of the user.}
	\item \mykey{user}{unique anonymized identifier of the user submitting the job. Users can submit jobs under multiple accounts, if they belong to multiple projects, so the same user should exhibit different behaviors based on the account used.}
	\item \mykey{partition}{name of the partition in which the job was submitted, among \texttt{\{htc, htc\_daemon, gpu, \dots\}}}, see Section~\ref{sec:partition}.
	\item \mykey{submit}{time at which the job was submitted. Formatted as Unix timestamp (start of the dataset on January 1st 2025 at 00:00 in GMT+0100 timezone corresponds to a timestamp of 1735686000). Jobs are sorted by this field.}
	\item \mykey{eligible}{time at which the job was eligible to be scheduled. Some jobs require to wait until a prescribed date or an event (e.g., another job termination) before being considered to be scheduled.}
	\item \mykey{start}{time at which the job is started.}
	\item \mykey{end}{time at which the job is terminated }
	\item \mykey{timelimit}{duration requested by the job, after which it will be terminated, in seconds.}
	\item \mykey{elapsed}{duration between \texttt{start} and \texttt{end}, in seconds.}
	\item \mykey{allocmem}{amount of memory required by the job and allocated to it, in megabytes.}
	\item \mykey{maxrss}{the maximum resident set size, or amount of memory consumed by the job at any given time, in megabytes. If a job fails because it required more memory, this number may still be below the threshold of \texttt{allocmem} as the last update of this field may have occurred before the threshold was hit. See Section~\ref{sec:caveatmem} for more details.

}
	\item \mykey{alloccpus}{number of CPUs, or virtual cores required by the job and allocated to it.}
	\item \mykey{totalcpu}{total CPU time used by the job, in seconds. For instance, a job using two CPUs at all time will have a \texttt{totalcpu} time equal to twice the \texttt{elapsed} time.}
	\item \mykey{allocgpus}{number of GPUs required by the job and allocated to it}
	\item \mykey{maxdiskwrite}{number of bytes written to disk. This is the only parameter that has not been actively monitored during the data gathering period, hence potential inconsistencies could not have been manually detected.}
	\item \mykey{el9}{custom field added to flag the jobs submitted on a machine running under  Red Hat Enterprise Linux 9. For such jobs, because of software issues, the reported \texttt{maxrss} value includes shared anonymous cache used for file copy so can be vastly overestimated, although this has no impact on which jobs are considered out-of-memory and killed, see Section~\ref{sec:caveatmem}. Equals either 0 or 1.}
	\item \mykey{hostnames}{anonymized identifiers of the machines on which the job is run. The only jobs running on multiple machines are those on the \texttt{hpc} partition.}
	\item \mykey{nsteps}{number of steps in which the job is split. Each can be allocated multiple CPUs, may be executed on distinct nodes, and parallel execution may be allowed by the user.}
	\end{itemize}
	
Note that several field can take a \texttt{NULL} value, such as the \texttt{maxrss} value of a job cancelled before its start.
Additionally, in order to improve reproducibility, the repository contains a Jupyter Notebook file analyzing the data and producing all statistics and plots presented in this paper.

\section{Description of the scheduling environment}

\label{sec:cluster}

\subsection{Partitions and jobs limits}
\label{sec:partition}
Machines are distributed among partitions depending on their type and usage. Most partitions allow job timelimits up to 7 days and restrict jobs to use a single machine.
Users can submit a job in one partition among:
\begin{itemize}
	\item \mykey{htc}{the main partition, by resources or by number of jobs, which are often submitted in batches.}
	\item \mykey{htc\_interactive}{allows interactive sessions so users can test their jobs before using the main partition.}
	\item \mykey{htc\_daemon}{dedicated to non-intensive jobs (also called "pilot jobs") that can last up to 90 days and manage others jobs. This is a separated partition so all long jobs are gathered on a single machine, facilitating maintenance.}
	\item \mykey{htc\_highmem}{partition dedicated to a single machine with high memory capacity.}
	\item \mykey{flash}{partition dedicated to short jobs ($<1$ hour).}
	\item \mykey{hpc}{partition allowing jobs to use multiple machines.}
	\item \mykey{gpu}{parition offering GPU hardware.}
	\item \mykey{gpu\_interactive}{allows interactive sessions using GPUs.}
\end{itemize}

The partitions \texttt{flash} and \texttt{htc\_daemon} share a single machine, and the repartition of machines among partition is depicted in Table~\ref{tab:partition}.

\begin{table}[htbp]
\centering
\caption{Repartition of resources among partitions.}
\label{tab:partition}
\begin{tabular}{cccccc}
	\hline
	\textbf{Machine ids} & \textbf{Partition} & \textbf{Nodes} & ~\textbf{CPUs} & ~\textbf{Memory} & ~\textbf{GPUs} \\
	\hline
	a03 & flash, htc\_daemon & 1 & 64 & 192 GB & 0 \\
	g[04-19] & gpu & 16 & 364 & 4426 GB & 64 \\
	g[01-02] & gpu\_interactive & 2 & 44 & 1738  GB& 8 \\
	p[01-16] & hpc & 16 & 512 & 2048 GB & 0 \\
	a[04-192],b[01-56],d[01-28] & htc & 273 & 21,952 & 93,996 GB & 0 \\
	m01 & htc\_highmem & 1 & 40 & 1546 GB & 0 \\
	a[01-02] & htc\_interactive & 2 & 128 & 384 GB & 0 \\
	\hline
\end{tabular}
\end{table}

As discussed in Section~\ref{sec:usage}, a memory overbooking mechanism is dedicated to compensate users overestimation: the \texttt{htc} workers declare more memory than they physically have, allowing more jobs to be scheduled. The manually adjusted overbooking amount averages around 30\% over the time period. It is therefore possible that few jobs respecting their memory demand get killed because of the overbooking system, in which case they are automatically rescheduled.

\subsection{Associations and limits}

\label{sec:caveatmem}

An \texttt{account} represents a group of individual \texttt{users}, where users
belong to at least one account. The combination of an account and a user is
called an \emph{association}. Two association corresponding to the same user can then have distinct rights and limits.
In the cluster, every account must always respect two limits:
\begin{itemize}
	\item \mykey{grpcpus}{total number of CPUs that can be concurrently used by associations member of this account ({2000} to {9600} CPUs).}
	\item \mykey{grpcmem}{total amount of memory that can be concurrently used by associations member of this account ({13} to {35} TB).}
\end{itemize}

When \texttt{grpcpus} or \texttt{grpmem} is attained, new jobs will stay pending until enough jobs from this account are terminated.
Additional limits to accounts, users or associations may be manually added temporarily. In particular, a \texttt{maxjobs} constraint may be set to momentarily limit running jobs of a problematic entity.

In addition, associations or accounts are associated to a fixed \texttt{share} value, influencing the priority of their jobs.

\subsection{Scheduling and resource allocation}

We include in this section the configuration details in order to improve the reproducibility of the associated data.
The scheduler uses the SLURM software, and associates a priority to each job submitted. 
This priority is determined through Slurm's Multifactor Priority Plugin\footnote{\url{https://slurm.schedmd.com/priority\_multifactor.html}} and depends on multiple factors. Pending jobs priority increase over time until \texttt{PriorityMaxAge}. The configuration is as follows: 

\begin{sloppypar}
\texttt{
PriorityDecayHalfLife   = 7-00:00:00;
PriorityCalcPeriod      = 00:05:00;
PriorityFavorSmall      = no;
PriorityMaxAge          = 4-00:00:00;
PriorityType            = priority/multifactor;
PriorityUsageResetPeriod = NONE;
PriorityWeightAge       = 1500;
PriorityWeightAssoc     = 0;
PriorityWeightFairShare = 1000.
}
\end{sloppypar}

 At every significant event (job submission or termination), jobs on the top of the priority queue are considered for execution. Furthermore, at every regular interval (30s), a backfilling procedure is performed to attempt at scheduling lower priority jobs if they would no delay future planned jobs. Note that jobs are always allocated the resources they request. The relevant settings are the following.

\begin{sloppypar}
\texttt{
SchedulerType           = sched/backfill;
SchedulerParameters     = pack\_serial\_at\_end,
                          max\_rpc\_cnt=400,
                          sched\_min\_interval=50000,
                          sched\_max\_job\_start=300,
                          batch\_sched\_delay=20,
                          bf\_resolution=600,
                          bf\_min\_prio\_reserve=2000,
                          bf\_min\_age\_reserve=600,
                          bf\_max\_job\_test=10000,
                          bf\_continue;
SchedulerTimeSlice      = 60 sec.
}
\end{sloppypar}

\subsection{Jobs in execution}

Running jobs are bound to resources via \texttt{cgroups} and do not share them nor are preempted, so a job has the full ownership on its allocated CPUs during its execution. It cannot use other CPUs, and if the time limit or allocated memory is reached, the job gets killed. As nodes are shared, jobs can be impacted by others because for instance of I/O contention.

\section{Limitations of the dataset}
\label{sec:caveats}

\subsubsection{Memory peak vs. usage}
The dataset does not report the memory utilization of a job throughout its execution but only the peak memory used, in the field \texttt{maxrss}. So a job achieving its peak for a small fraction of time is not distinguished from a job often using fully its memory for a prolonged duration.

\subsubsection{MaxRSS and Red Hat Enterprise Linux 9}
Because of system updates, when migrating to Red Hat Enterprise Linux 9, the memory usage reported through the \texttt{maxrss} field includes shared anonymous cache used for file copy. The consequence being that this field has a value vastly overestimated, often equal to the job allocated memory. This had no impact on the job processing (such jobs are not killed erroneously), but compromised the reporting. Such jobs are distinguished thanks to the field \texttt{el9}. The migration ended in September, so all jobs after this date and some jobs before are concerned, totaling $14.6$M jobs.

\subsubsection{MaxRSS and number of steps}
Jobs are composed of \emph{steps}, the number of which being indicated by the field \texttt{nsteps}. A job with a single step may run on multiple CPUs but on a single node. A job with multiple steps may execute them in parallel, sequentially, or a mix of both. The field \texttt{maxrss} reports the maximum value of \texttt{maxrss} over all steps. Therefore, the value is correct for multi-step sequential jobs. For multi-step parallel jobs, the actual memory peak can be as high as \texttt{max(allocmem,maxrss$\cdot$nsteps)}. Unfortunately, it is not possible to guarantee whether a multi-step job is parallel or sequential. The comparison of the job duration and the total CPU time divided by the number of CPUs allocated is a good indicator.

\subsubsection{I/O impact}
The amount of I/O performed can significantly impact the performance of a job, especially as multiple filesystems are shared and distributed. The dataset therefore includes the \texttt{maxdiskwrite} field collected through SLURM tools, but this value has never been monitored while in production.

\subsubsection{Negative waiting time}
Because of a reporting bug, 116 jobs have a start time anterior to their eligible time, by up to one hour.

\subsubsection{Completed and Failed state}
A job which has not been killed has a reported state either \emph{completed} or \emph{failed}. It is determined based on the exit code of the bash script submitted. Therefore, scripts with improper exit code handling can lead to inconsistent reported state.

\section{Statistical analysis of the dataset}
\label{sec:analysis}

\subsection{General description}
\label{sec:introstat}

The dataset contains 44,749,836 jobs among which 44,742,965 were submitted during 2024.
During the time window considered, 945 unique users have submitted a job, and they are grouped in 101 accounts. As a user may belong to several accounts, this amounts to 1053 distinct associations of a user and an account.

Almost all jobs are submitted to the \texttt{htc} partition (99.5\%), followed by \texttt{hpc} (0.3\%) and \texttt{gpu} (0.1\%). The predominance of \texttt{htc} usage is slightly decreased when focusing on the global share of allocated CPU time (96.6\%) or memory (95.9\%) as other jobs need more resources on average. One account (account035) and in particular one user of this account (user0137) submitted a significant portion of the jobs in the dataset, 27.9\%. Therefore, for several descriptors where this account introduces a large bias, we report statistics with and without this account, which amount to 35.7\% of the jobs submitted, and used 18.7\% of the CPU time and 13.2\% of memory allocated.

Most jobs terminate in the \texttt{completed} state (82.9\%), followed by \texttt{fail} (12.1\%), \texttt{cancelled} (3.4\%), \texttt{timeout} (1.22\%) and \texttt{out\_of\_memory} (0.4\%). Note that a significant part of failed jobs terminate in a very short time (median time of 38s), a typical reason being file or library dependency issues, which happen at the start of the execution.

Jobs submitted to the \texttt{hpc} partition are more likely to be composed of several \emph{steps} (18.3\%) than other jobs (1.9\%), and such jobs are predominantly composed of exactly two steps.

Most jobs are executed on a single core (89.9\%). The following most requested CPU numbers are small even numbers: 2, 4, 6, 8, 10, 12 CPUs are respectively requested by 3.2, 4.2, 1.4, 0.7, 0.3 and 2.5 percent of jobs.

\begin{figure}[tbp!]
	\centering
	\includegraphics[width=.49\textwidth]{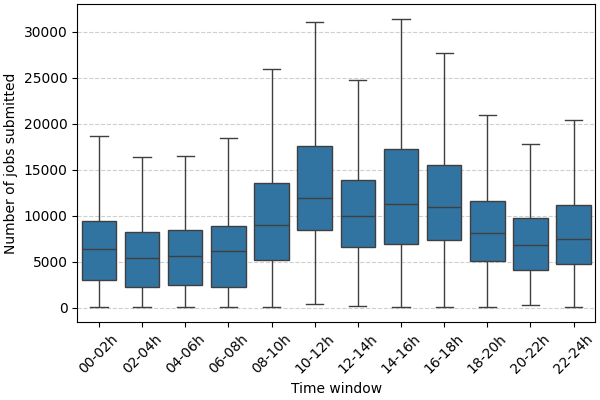}
	\includegraphics[width=.49\textwidth]{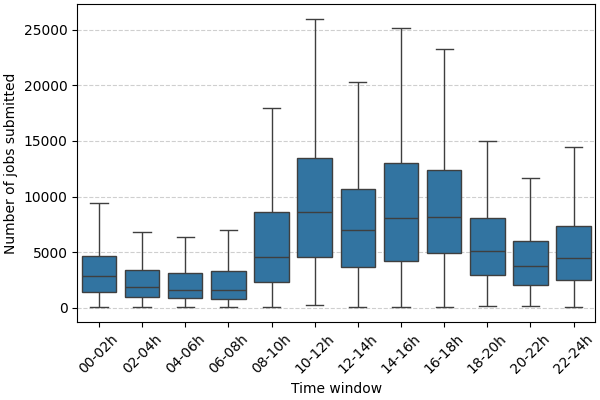}
	\caption{Statistical description of the number of jobs submitted per time of the day (median, quartiles and $1.5\cdot$IQR whiskers without outliers). Right plot without \texttt{account035}.}
	\label{fig:submithour}
\end{figure}

\begin{figure}[tbp!]
	\centering
	\includegraphics[width=.49\textwidth]{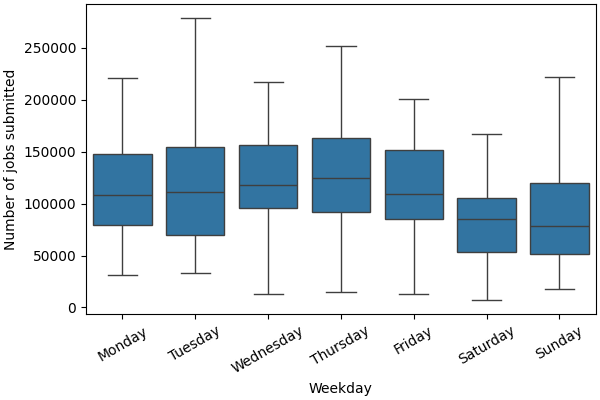}
	\includegraphics[width=.49\textwidth]{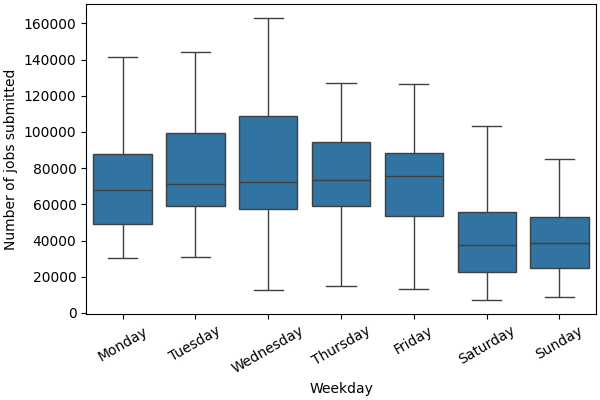}
	\caption{Statistical description of the number of jobs submitted per day of the week (median, quartiles and $1.5\cdot$IQR whiskers without outliers). Right plot without \texttt{account035}.}
	\label{fig:jobsbyweekday}
\end{figure}

\subsection{Submission time statistics}

\begin{figure}[tbp!]
	\centering
	\includegraphics[width=\textwidth]{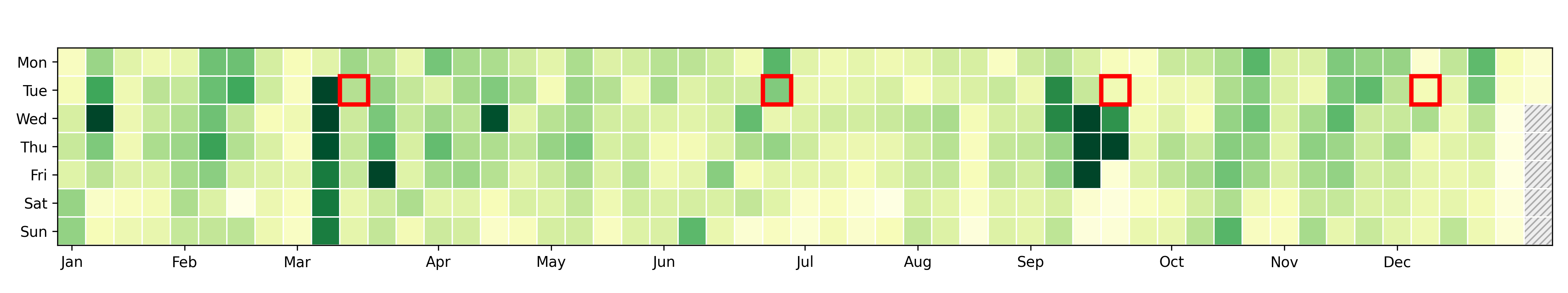}
	\includegraphics[width=\textwidth]{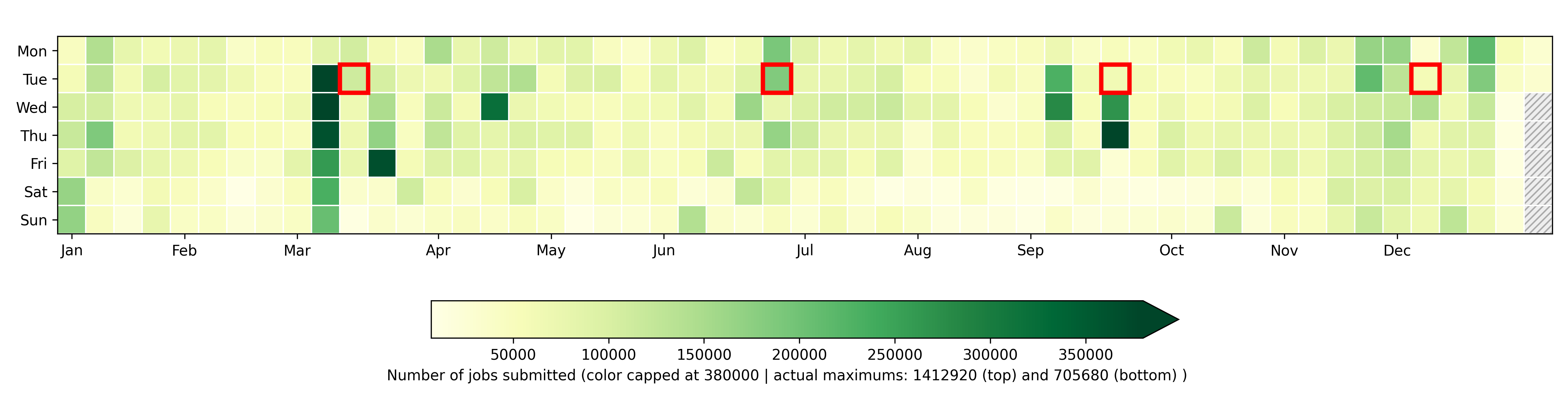}
	\caption{Calendar of job submissions with maintenance days highlighted, color scale capped at the 98th percentile of the data (bottom without \texttt{account035}).}
	\label{fig:submitcal}
\end{figure}

Many jobs are submitted either manually by users, or through workflow managers requiring human intervention, hence the rate of submission depends on the time of day (Figure~\ref{fig:submithour}), the day of the week (Figure~\ref{fig:jobsbyweekday}) or the day of the year (Figure~\ref{fig:submitcal}). Office hours and weekdays witness higher submission rates, although the difference is not dramatic. There are also several high-demand days over the year, and traditional holiday periods are associated with low activity.

\subsection{Waiting time }

\begin{figure}[tbp!]
	\centering
	\includegraphics[scale=0.4]{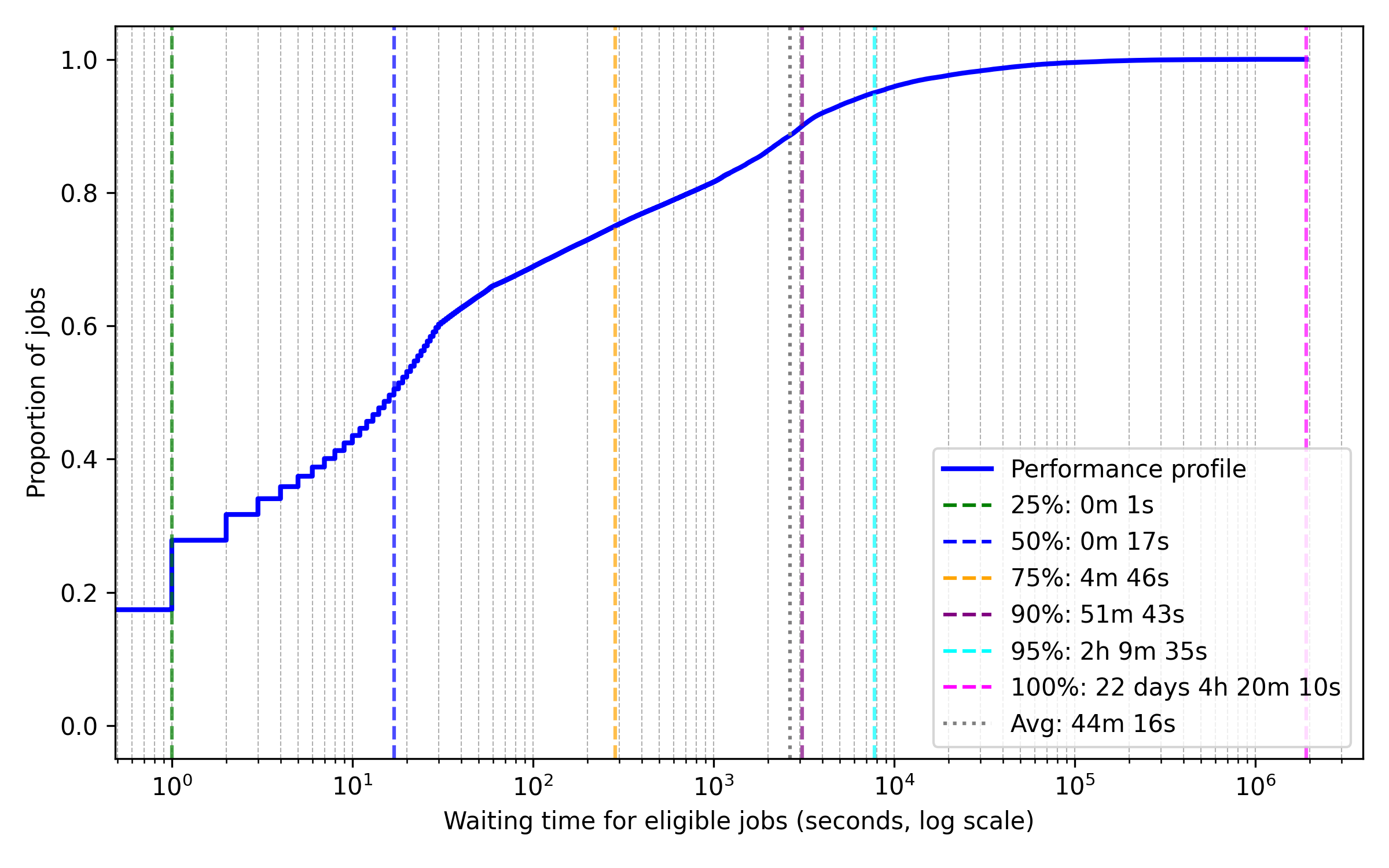}
	\caption{Performance profile of the waiting time of eligible jobs. A point $(x,y)$ means that a proportion $y$ of all jobs have a waiting time at most $x$ (high-left is best).}
	\label{fig:cdfwaiting}
\end{figure}

The vast majority of jobs become eligible in negligible time ($<4$s for 99\% of jobs) hence we focus here on the duration between the moments when jobs become eligible and start running. This duration averages 44min but, as seen on Figure~\ref{fig:cdfwaiting}, half of the jobs wait less than 17s. Longer waiting times occur for instance when the cluster is saturated, user limits are reached or jobs request rare resources (e.g., all CPUs of a machine). A discontinuity can be observed at 30s and 1min: the scheduler performs backfilling passes every 30s, so jobs able to be scheduled via the first backfilling pass have a waiting time drawn uniformly within 30s.

\subsection{Jobs accuracy and efficiency}

\begin{figure}[tbp!]
	\centering
	\includegraphics[width=.49\textwidth]{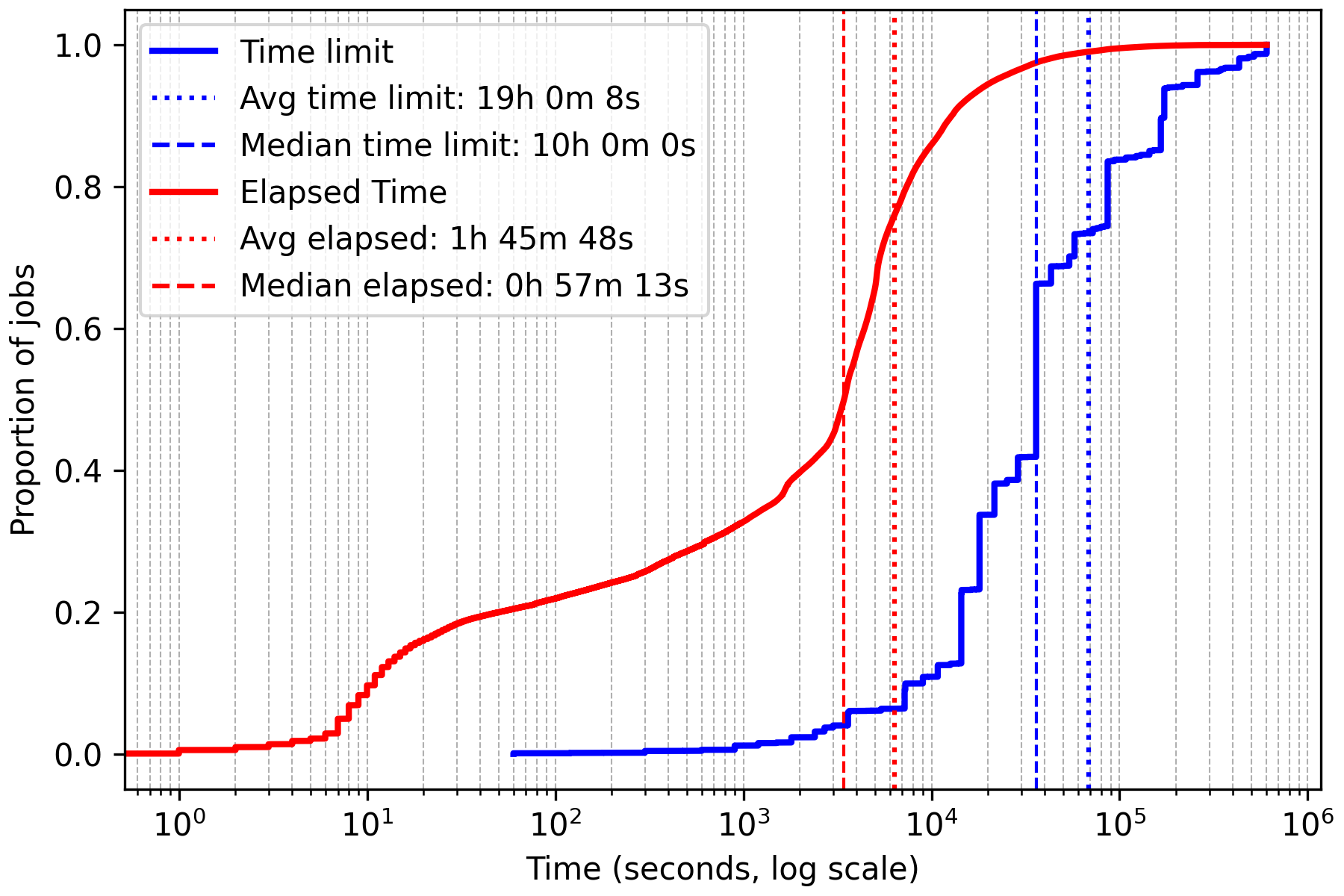}
	\includegraphics[width=.49\textwidth]{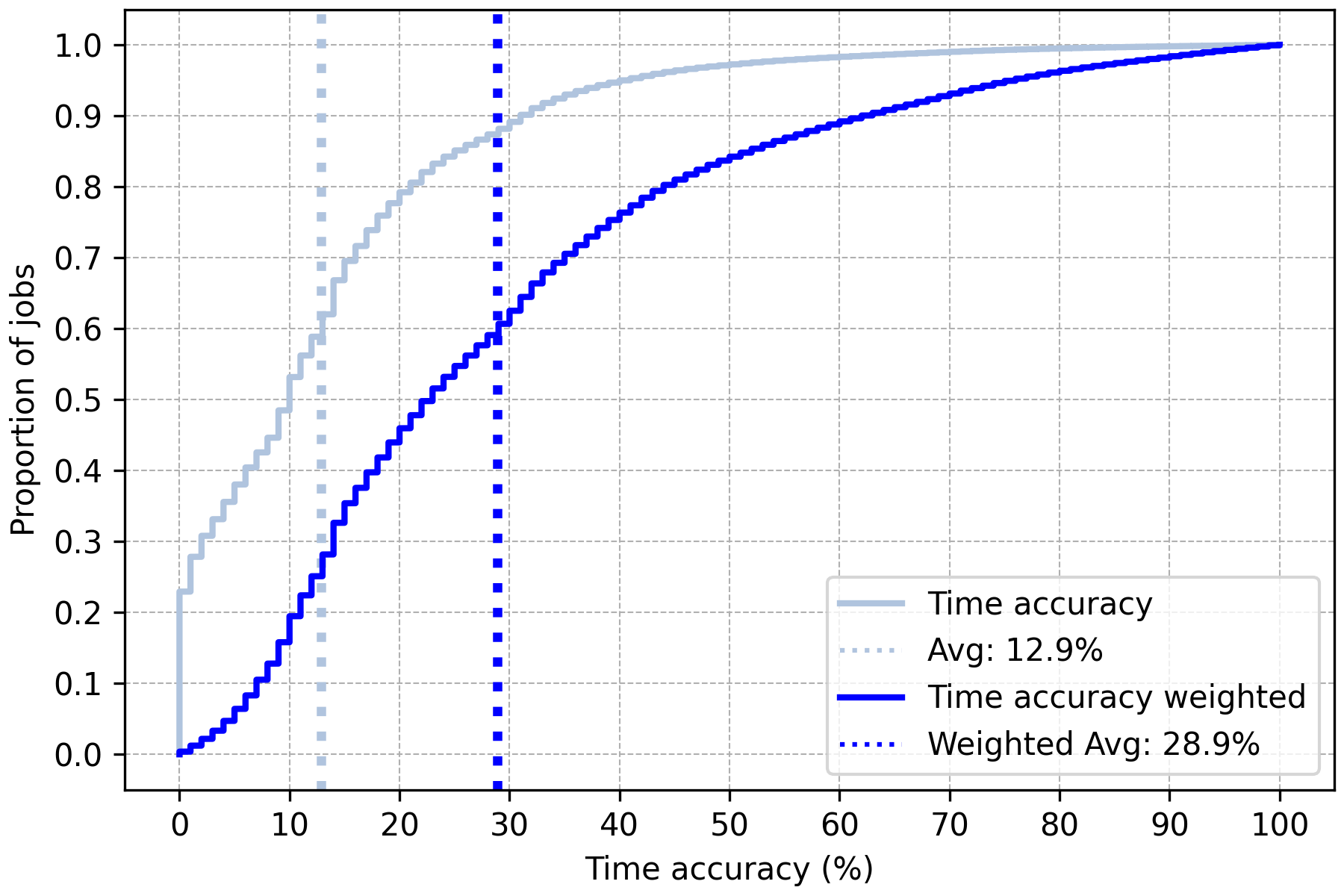}
	\caption{Performance profile of the requested time limit and computation time (left) and of the percentage of requested time used (right).}
	\label{fig:cdftimeeff}
\end{figure}

The following statistics focus on completed (not failed) jobs and exclude the partition \texttt{htc\_daemon}, composed of long pilot jobs not aiming at efficiency.

Users vastly overestimate the duration needed to terminate their jobs, the median usage of allocated time for completed jobs is equal to around 10\% of the time limit requested, see Figure~\ref{fig:cdftimeeff}. When weighting each job by its duration, the accuracy statistics improve, the median surpassing 20\%. This can be explained as low-efficient jobs terminating prematurely now have minimal impact, and long jobs are more studied by users.

\begin{figure}[tbp!]
	\centering
	\includegraphics[width=\textwidth]{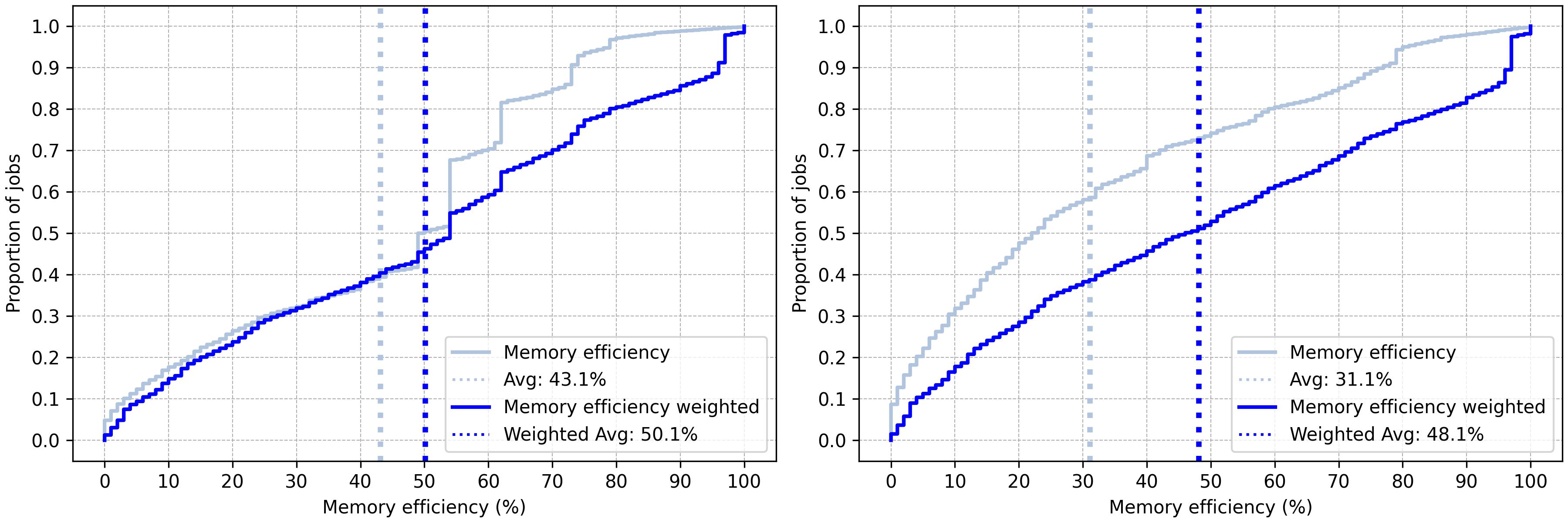}
	\caption{Performance profile of the percentage of memory used with (left) and without (right) \texttt{account035}  (lower-right is best).}
	\label{fig:cdfmemeff}
\end{figure}

Figure~\ref{fig:cdfmemeff} shows the repartition of memory efficiency for completed jobs that ran on non-el9 machines (see Section~\ref{sec:caveats}). This amounts to $21.8$M jobs, and $12.1$M jobs without the most predominant account. Both cumulative distribution (by number of jobs) and weighted cumulative distribution are represented. The weight is proportional to the duration of the jobs and the memory allocated, as long jobs and memory-intensive jobs have more impact on the overall efficiency. The bias introduced by a single account (see Section~\ref{sec:introstat}) is highly noticeable here: large peaks of jobs having the exact same efficiency are visible especially on the unweighted left plot. Note that the weighted efficiency presents better statistics that the non-weighted counterpart. This means that more demanding jobs (resourcewise) are more efficient that jobs running for a short time or using less memory. It can be explained by low-memory jobs often requesting a minimum threshold value, while high-memory jobs are more studied and make requests closer to their needs.

\begin{figure}[tbp!]
	\centering
	\includegraphics[width=\textwidth]{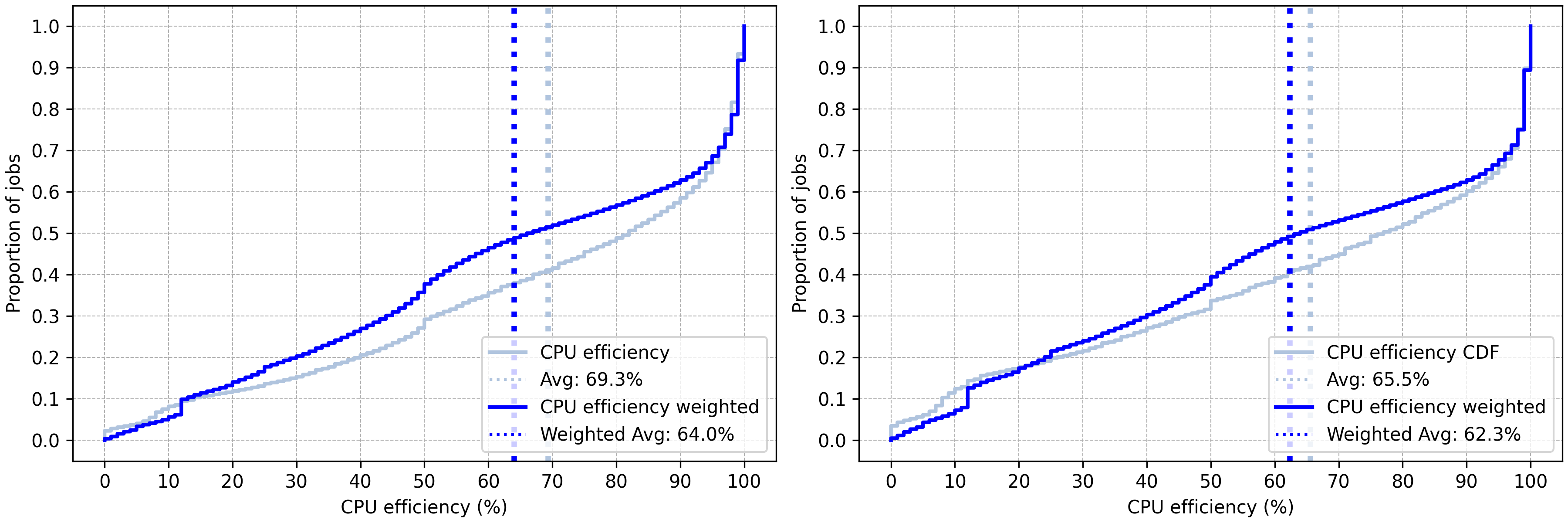}
	\caption{Performance profile of CPU efficiency with (left) and without (right) \texttt{account035}  (lower-right is best).}
	\label{fig:cdfcpueff}
\end{figure}

Figure~\ref{fig:cdfcpueff} show the same analysis regarding CPU time efficiency (the total CPU time used divided by the job duration times the number of CPUs allocated), which can be affected by e.g., intensive I/O usage or suboptimal use of multiple cores. The jumps observed at 12.5\%, 25\% and 50\% can then be explained by jobs requesting 8, 4 and 2 CPUs while using one of them most of the time.
We can see that, contrarily to memory, the weighted efficiency is worse than the non-weighted. Indeed, jobs requesting more CPUs or lasting longer are more encline to idle time, because of parallelism or I/O contention. The effect of the most used account is here not as significant as regarding memory efficiency.

 \bibliographystyle{splncs04}
 \bibliography{workload}

@article{swim,
  title={Interactive Analytical Processing in Big Data Systems: A Cross-Industry Study of MapReduce Workloads},
  author={Chen, Yanpei and Alspaugh, Sara and Katz, Randy},
  journal={Proceedings of the VLDB Endowment},
  volume={5},
  number={12},
  year={2012},
  url={https://github.com/SWIMProjectUCB/SWIM}
}

@inproceedings{swf,
  title={Benchmarks and standards for the evaluation of parallel job schedulers},
  author={Chapin, Steve J and Cirne, Walfredo and Feitelson, Dror G and Jones, James Patton and Leutenegger, Scott T and Schwiegelshohn, Uwe and Smith, Warren and Talby, David},
  booktitle={Job Scheduling Strategies for Parallel Processing: IPPS/SPDP’99Workshop, JSSPP’99 San Juan, Puerto Rico, April 16, 1999 Proceedings 5},
  pages={67--90},
  year={1999},
  organization={Springer},
  url={https://www.cs.huji.ac.il/labs/parallel/workload/swf.html}
}

@article{pwa,
  title={Experience with using the parallel workloads archive},
  author={Feitelson, Dror G and Tsafrir, Dan and Krakov, David},
  journal={Journal of Parallel and Distributed Computing},
  volume={74},
  number={10},
  pages={2967--2982},
  year={2014},
  publisher={Elsevier},
  url={www.cs.huji.ac.il/labs/parallel/workload/}
}

@inproceedings{obfuscation,
  title={Obfuscatory obscanturism: making workload traces of commercially-sensitive systems safe to release},
  author={Reiss, Charles and Wilkes, John and Hellerstein, Joseph L},
  booktitle={2012 IEEE Network Operations and Management Symposium},
  pages={1279--1286},
  year={2012},
  organization={IEEE}
}

@inproceedings{gwa,
  title={How are real grids used? the analysis of four grid traces and its implications},
  author={Iosup, Alexandru and Dumitrescu, Catalin and Epema, Dick and Li, Hui and Wolters, Lex},
  booktitle={2006 7th IEEE/ACM International Conference on Grid Computing},
  pages={262--269},
  year={2006},
  organization={IEEE},
  url={https://atlarge-research.com/gwa.html}
}

@inproceedings{google,
  title={Borg: the next generation},
  author={Tirmazi, Muhammad and Barker, Adam and Deng, Nan and Haque, Md E and Qin, Zhijing Gene and Hand, Steven and Harchol-Balter, Mor and Wilkes, John},
  booktitle={Proceedings of the fifteenth European conference on computer systems},
  pages={1--14},
  year={2020},
  url={https://github.com/google/cluster-data/tree/master}
}

@inproceedings{alibaba,
  title={GPU-Disaggregated Serving for Deep Learning Recommendation Models at Scale},
  author={Yang, Lingyun and Wang, Yongchen and Yu, Yinghao and Weng, Qizhen and Dong, Jianbo and Liu, Kan and Zhang, Chi and Zi, Yanyi and Li, Hao and Zhang, Zechao and others},
  booktitle={22nd USENIX Symposium on Networked Systems Design and Implementation (NSDI 25)},
  pages={847--863},
  year={2025},
  url={https://github.com/alibaba/clusterdata}
}

@article{fta,
  title={The Failure Trace Archive: Enabling the comparison of failure measurements and models of distributed systems},
  author={Javadi, Bahman and Kondo, Derrick and Iosup, Alexandru and Epema, Dick},
  journal={Journal of Parallel and Distributed Computing},
  volume={73},
  number={8},
  pages={1208--1223},
  year={2013},
  publisher={Elsevier},
  url={http://fta.scem.westernsydney.edu.au/}
}

@inproceedings{slurm,
  title={Architecture of the slurm workload manager},
  author={Jette, Morris A and Wickberg, Tim},
  booktitle={Workshop on Job Scheduling Strategies for Parallel Processing},
  pages={3--23},
  year={2023},
  organization={Springer}
}

@inproceedings{vasconcelos2023optimal,
  title={Optimal sizing of a globally distributed low carbon cloud federation},
  author={Vasconcelos, Miguel and Cordeiro, Daniel and Da Costa, Georges and Dufoss{\'e}, Fanny and Nicod, Jean-Marc and Rehn-Sonigo, Veronika},
  booktitle={2023 IEEE/ACM 23rd International Symposium on Cluster, Cloud and Internet Computing (CCGrid)},
  pages={203--215},
  year={2023},
  organization={IEEE}
}

@inproceedings{amvrosiadis2018diversity,
  title={On the diversity of cluster workloads and its impact on research results},
  author={Amvrosiadis, George and Park, Jun Woo and Ganger, Gregory R and Gibson, Garth A and Baseman, Elisabeth and DeBardeleben, Nathan},
  booktitle={2018 USENIX Annual Technical Conference (USENIX ATC 18)},
  pages={533--546},
  year={2018}
}

@article{haddad2021combined,
  title={Combined IT and power supply infrastructure sizing for standalone green data centers},
  author={Haddad, Marwa and Da Costa, Georges and Nicod, Jean-Marc and P{\'e}ra, Marie-C{\'e}cile and Pierson, Jean-Marc and Rehn-Sonigo, Veronika and Stolf, Patricia and Varnier, Christophe},
  journal={Sustainable Computing: Informatics and Systems},
  volume={30},
  pages={100505},
  year={2021},
  publisher={Elsevier}
}

@inproceedings{villebonnet2016energy,
  title={Energy aware dynamic provisioning for heterogeneous data centers},
  author={Villebonnet, Violaine and Da Costa, Georges and Lef{\`e}vre, Laurent and Pierson, Jean-Marc and Stolf, Patricia},
  booktitle={2016 28th international symposium on computer architecture and high performance computing (SBAC-PAD)},
  pages={206--213},
  year={2016},
  organization={IEEE}
}

@article{haddad2021stand,
  title={Stand-alone renewable power system scheduling for a green data center using integer linear programming},
  author={Haddad, Maroua and Nicod, Jean-Marc and P{\'e}ra, Marie-C{\'e}cile and Varnier, Christophe},
  journal={Journal of Scheduling},
  volume={24},
  number={5},
  pages={523--541},
  year={2021},
  publisher={Springer}
}

@inproceedings{benaissa2022standalone,
  title={Standalone data-center sizing combating the over-provisioning of the it and electrical parts},
  author={Benaissa, Manal and Da Costa, Georges and Nicod, Jean-Marc},
  booktitle={2022 International Symposium on Computer Architecture and High Performance Computing Workshops (SBAC-PADW)},
  pages={57--62},
  year={2022},
  organization={IEEE}
}

@inproceedings{vasconcelos2022indirect,
  title={Indirect network impact on the energy consumption in multi-clouds for follow-the-renewables approaches},
  author={Vasconcelos, Miguel Felipe Silva and Cordeiro, Daniel and Dufoss{\'e}, Fanny},
  booktitle={11th International Conference on Smart Cities and Green ICT Systems},
  pages={44--55},
  year={2022},
  organization={SCITEPRESS-Science and Technology Publications}
}

@article{marconi,
  title={M100 exadata: a data collection campaign on the cineca’s marconi100 tier-0 supercomputer},
  author={Borghesi, Andrea and Di Santi, Carmine and Molan, Martin and Ardebili, Mohsen Seyedkazemi and Mauri, Alessio and Guarrasi, Massimiliano and Galetti, Daniela and Cestari, Mirko and Barchi, Francesco and Benini, Luca and others},
  journal={Scientific Data},
  volume={10},
  number={1},
  pages={288},
  year={2023},
  publisher={Nature Publishing Group UK London}
}

@inproceedings{cendrier2025green,
  title={Green scheduling on the edge},
  author={Cendrier, Joachim and Wijayawardana, Rajini and Benoit, Anne and Robert, Yves and Vivien, Fr{\'e}d{\'e}ric and A. Chien, Andrew},
  booktitle={European Conference on Parallel Processing},
  pages={380--394},
  year={2025},
  organization={Springer}
}

\end{document}